\documentclass{article}
\usepackage{color}
\usepackage{verbatim}
\usepackage{amsmath, amssymb}       
\usepackage{fancybox,subfigure}
\usepackage{hyperref}
\usepackage[T1]{fontenc}
\newlength{\querylen}
\usepackage{tikz}
\usepackage{tikz-qtree}
\usetikzlibrary{bayesnet}
\usepackage{natbib}
\usepackage[plain,noend]{algorithm2e}

\makeatletter
\renewcommand{\algocf@captiontext}[2]{#1\algocf@typo. \AlCapFnt{}#2} 
\def\@algocf@capt@plain{top}
\renewcommand{\algocf@makecaption}[2]{%
  \addtolength{\hsize}{\algomargin}%
  \sbox\@tempboxa{\algocf@captiontext{#1}{#2}}%
  \ifdim\wd\@tempboxa >\hsize
    \hskip .5\algomargin%
    \parbox[t]{\hsize}{\algocf@captiontext{#1}{#2}}
  \else%
    \global\@minipagefalse%
    \hbox to\hsize{\box\@tempboxa}
  \fi%
  \addtolength{\hsize}{-\algomargin}%
}
\makeatother

\newcommand\Nor{\mathcal{N}}

\usepackage{titlesec} 
\titleformat{\subsection}[runin]
{\normalfont\bfseries}{\thesubsection}{1em}{}

\begin{document}

%
%
%
%
\title{A note on MCMC for nested multilevel regression models via
  belief propagation}
 
\author{Omiros Papaspiliopoulos\footnote{Instituci\'o Catalana de Recerca i Estudis Avan\c{c}ats,  Ramon
  Trias Fargas 25-27, Barcelona
  08005
  \href{mailto:omiros.papaspiliopoulos@upf.edu}{omiros.papaspiliopoulos@upf.edu}
}\; and 
 Giacomo Zanella\footnote{
Department of Decision Sciences, BIDSA and IGIER, Bocconi University, via Roentgen 1, 20136
Milan, Italy.
\href{mailto:giacomo.zanella@unibocconi.it}{giacomo.zanella@unibocconi.it}.
GZ supported by the European Research Council (ERC) through StG ``N-BNP'' 306406.
}
}

\maketitle
\begin{abstract}
In the quest for scalable Bayesian computational algorithms we need to exploit
the full potential of existing methodologies. In this note we show that the posterior distribution of the regression
parameters conditionally on covariance hyperparameters in a large
family of multilevel regression models is a
high-dimensional Gaussian that can be sampled exactly (as well as marginalized) using belief
propagation at a cost that scales linearly in the number of parameters and
data.
We derive an algorithm that works efficiently even for conditionally singular Gaussian distributions, e.g., when there are linear constraints between the parameters at different
levels. We provide a synthesis of similar ideas that have been
suggested in the literature before. 
\end{abstract}

\section{Belief propagation}
\label{sec:background}

Graphical models are now mainstream for modelling high-dimensional
vectors of dependent random variables and for doing efficient
computations with the
joint density thereof, such as marginalisations and maximisations. 
Among others, two algorithmic paradigms have been established
for scalable computations in graphs.
More popular
within statistics is the junction tree algorithm, see for example
Chapter 6 of \cite{expert}, and  more popular within
machine learning is belief propagation, see for example Section 8.4
in \cite{bishop}. The connections between the two paradigms are
understood, see for example Chapter 2 of \cite{jordan}. For
the class of models we consider in this article the two paradigms are
essentially equivalent, so we will use the belief propagation
formulation, which we find more intuitive in our context and we
describe below. 
The starting point of belief propagation is a factorisation of the joint density of a vector
of random variables. (In Section \ref{sec:bp-nested} we work with a
broader framework that starts with a factorisation of the joint law of
the variables.) The set of variables and factors are then
represented by a bipartite graph known as the factor
graph. There are two sets of nodes: variable nodes (one for each of
the random variables) and factor nodes (one for each of the
factors in the density factorisation). The edges in the graph connect each factor node to
the variable nodes that correspond to those variables involved in the
corresponding factor in the factorisation. Details on this
construction can be found for example in Section 8.4 of
\cite{bishop}. 

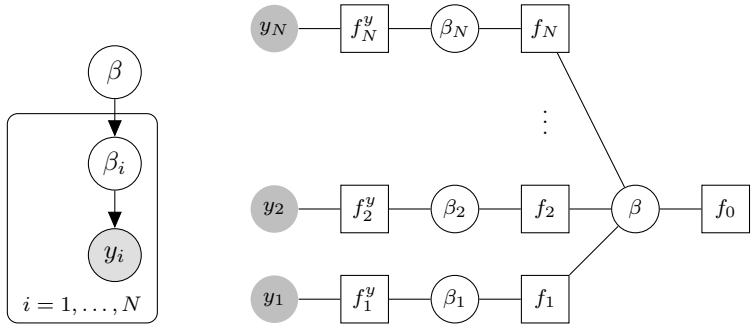
\begin{figure}[htbp]
\centering
  \tikz{
     \node[obs] (y) {$y_{i}$};%
     \node[latent,above=of y,yshift=-0.5cm] (eta) {$\beta_{i}$}; %
     \node[latent,above=of eta,yshift=-0.5cm] (gamma) {$\beta$}; %
     \plate [inner sep=0.2cm,xshift=0cm,yshift=0.1cm] {plate1} {(y)(eta)} {$i=1,\dots,N$}; %
     \edge {eta} {y}
     \edge {gamma} {eta}
}   
\hspace{10mm}
\begin{tikzpicture}[ > = triangle 45, scale = .6]
\tikzstyle{every node} = [font = \footnotesize]
\tikzstyle{vertexy} = [circle, fill = black!25, minimum size = 18pt, inner sep = 1pt]
\tikzstyle{vertexx} = [circle, draw, solid, minimum size = 18pt, inner
sep = 1pt]
\tikzstyle{vertexf} = [rectangle, draw, solid, minimum size = 18pt, inner sep = 1pt]
\tikzstyle{empty} = [circle, fill = white, minimum size = 18pt, inner sep = 1pt]
\node[vertexf] (f0) at (2, 0) {$f_0$};
\node[vertexx] (x) at (0, 0) {$\beta$};
\node[vertexf] (f1) at (-2, -2) {$f_1$};
\node[vertexf] (f2) at (-2, 0) {$f_2$};
\node[empty] (dots) at (-2, 2.1) {$\vdots$};
\node[vertexf] (f3) at (-2, 4) {$f_N$};
\node[vertexx] (x1) at (-4, -2) {$\beta_1$};
\node[vertexx] (x2) at (-4, 0) {$\beta_2$};
\node[vertexx] (x3) at (-4, 4) {$\beta_N$};
\node[vertexf] (f4) at (-6, -2) {{$f^y_{1}$}};
\node[vertexf] (f5) at (-6, 0) {{$f^y_{2}$}};
\node[vertexf] (f6) at (-6, 4) {{$f^y_{N}$}};
\node[vertexy] (y1) at (-8, -2) {$y_1$};
\node[vertexy] (y2) at (-8, 0) {$y_2$};
\node[vertexy] (y3) at (-8, 4) {$y_N$};
\draw[-] (x) edge (f0);
\draw[-] (x) edge (f1);
\draw[-] (x) edge (f2);
\draw[-] (x) edge (f3);
\draw[-] (f1) edge (x1);
\draw[-] (f2) edge (x2);
\draw[-] (f3) edge (x3);
\draw[-] (x1) edge (f4);
\draw[-] (x2) edge (f5);
\draw[-] (x3) edge (f6);
\draw[-] (f4) edge (y1);
\draw[-] (f5) edge (y2);
\draw[-] (f6) edge (y3);
\end{tikzpicture}
        \caption{ {Two}-level hierarchical model as directed acyclic
          graph (left) and as a factor graph (right). The notation
          used for the factor nodes is explained in Section \ref{sec:bp-nested}.}
        \label{fig:fact_graph}
\end{figure}

An example is the two-level hierarchical model
\begin{equation}\label{eq:model_2_levels}
\begin{aligned}
\beta &  \sim \Nor(\mu_0,V_0)\\
\beta_{i} \mid \beta & \sim \Nor(\beta,\Sigma) \\
y_{i}  \mid \beta_i & \sim \Nor(X^y_{i}\beta_{i},\sigma_i^2 I)
\end{aligned}
\end{equation}
 where $I$ is the identity matrix whose dimensions
vary depending on the context; $\Nor(\mu,\Sigma)$ denotes the
Gaussian distribution, later $\Nor(x;\mu,\Sigma)$ will denote the corresponding Gaussian
density with argument $x$; $X_i^y$ is a matrix of covariates, one for
each $i$, where superscription by $y$ is explained in Section
\ref{sec:bp-nested} in the context of a generalisation of
\eqref{eq:model_2_levels}. 
 The last line of \eqref{eq:model_2_levels} defines a
linear regression model, $y_{ij} \sim \Nor\big((X^y_{ij})^T
\beta_i,\sigma_i^2\big)$, where $(X^y_{ij})^T$ is the $j$th row of matrix $X^y_i$.
The second line of the model pools the local regression coefficients for
each $i$ towards a global parameter vector $\beta$. The directed acyclic graph of this
model for given covariance and design matrices is shown in Figure
\ref{fig:fact_graph}(left). Each of the conditional densities in the
model specification corresponds to a factor and the associated factor
graph is shown in Figure \ref{fig:fact_graph}(right). 

Belief propagation involves exchange of messages
between variable and factor nodes.  Let $t$ denote a
variable node in the graph, $\beta_t$ the corresponding variable, $s$ a
neighbouring factor node, $f_s$ the corresponding factor in the
density factorisation, and
$ne(\cdot)$ be a function that takes as input a node index and returns
the set of neighbouring nodes in the graph.
By construction factor
nodes neighbour variable nodes only and variable nodes neighbour
factor nodes only.
Hence, $f_s$ is a function of $\beta_t$ and
$\beta_{ne(s)\backslash t}$. 
Messages exchanged between variable and factor nodes are functions of
the variable attached to the variable node {and are defined by the following equations:}
\begin{equation}
  \label{eq:messages}
\begin{aligned}
  m_{\beta_t \to f_s} (x) & = \prod_{k \in ne(t)\backslash s} m_{f_k \to
    \beta_t} (x) \\
m_{f_s \to \beta_t} (z_t) & = \int f_s(z_{ne(s)}) \prod_{j \in ne(s)\backslash t} m_{\beta_j
\to f_s}(z_j) d z_j. 
\end{aligned}
\end{equation}
In terms of notation used in this paper,  for sets $A$ and $B$,  $A\backslash B$ denotes the set
difference, we identify the variable index $t$ 
with the one-element set that contains it, for a set of indices
$A$, $\beta_A$ denotes the set of variables indexed by elements in $A$, $x$
and $z$ denote generic function arguments. 
If the factor graph is a tree, as for example in Figure \ref{fig:fact_graph}(right), and the integrals in \eqref{eq:messages} can be computed, then the system of equations defined by \eqref{eq:messages} can be efficiently solved with a forward-backward algorithm.
After choosing an arbitrary variable node as the root of
the tree, the direction from the leaves to the root is defined as
forward  and the reverse as backward. For a given node $t$, its
neighbour on its unique path towards the root will be called its
parent and denoted by $pa(t)$.  The algorithm is initialised by
setting  the messages at the variables nodes at the leaves (if any) equal to 1. The
forward pass of belief propagation {computes messages going}
from the leaves to the root and the backward pass computes messages  from the root to the leaves. 
{After two sweeps we obtain the set of messages solving
  \eqref{eq:messages}.} The resulting messages can then be used to compute the marginal density at any variable node $t$ as $p_{\beta_t}(x) = \prod_{s \in ne(t)} m_{f_s \to \beta_t}(x)$.
When some variables in the
definition of the graphical model have been observed, say $y_A$,  the
message passing steps that involve these variables keep them fixed at
their observed values and do not integrate them out.  Then,
after the forward and backward step the product of the messages at a
variable node $t$ is proportional to the conditional density
$p_{\beta_t|y_A}(x)$. 
{Although }belief propagation is typically developed for computing marginal
distributions and normalising constants, {it can also be used to
  simulate from conditional distributions. This is exploited in the
  following section, where in the backward pass simulation replaces
  message computation.}

\section{Bayesian nested multilevel models}
\label{sec:bp-nested}

\subsection{Model formulation} \label{sec:model_formulation}
The methods we develop are appropriate for $(k+1)$-level models with the
following nested structure
\begin{equation}\label{eq:nested_models}
\begin{aligned}
\beta \sim \Nor(\mu_0,V_0) \quad & \textrm{or}  \quad \beta \sim p(\beta)
\propto 1\\
\beta_i \mid \beta & \sim \Nor(X_i \beta,\Sigma) \\ 
\beta_{i_1\dots i_d} \mid \beta_{i_1\dots i_{d-1}} & \sim
\Nor({X_{i_1\dots i_{d}}} \beta_{i_1\dots i_{d-1}},\Sigma_{i_1\dots
  i_{d-1}}),  \quad d=2,\dots,k \\
y_{i_1\dots i_k} \mid \beta_{i_1\dots i_k} & \sim \Nor(X^y_{i_1\dots i_k}
\beta_{i_1\dots i_k}, \sigma_{i_1\dots i_k}^2I). 
\end{aligned}
\end{equation}
The level closest to the data will
be understood as the deepest one.
The last line of \eqref{eq:nested_models} is a linear regression model where we have
concatenated observations at the deepest level into the vector
$y_{i_1\dots i_k}${, like we did in \eqref{eq:model_2_levels}}. 
We allow for level-specific covariates that explain the variation of
regression coefficients at that level. Throughout the following
presentation there will be two sets of quantities associated to the
deepest level and that are subscripted in the same way. To distinguish
between them, we add the superscript $y$ to those that link to the
data; therefore, $X^y_{i_1\dots i_k}$ is the matrix of covariates used
to explain the dataset $y_{i_1\dots i_k}$, whereas $X_{i_1\dots i_k}$
is used to explain $\beta_{i_1\dots
  i_k}$. 
 We also allow for flat prior on
the hyperparameters, which yields a proper posterior under a condition
on the matrix $P_0$ defined in \eqref{eq:post-root} below.
The set of all observations $y_{i_1\dots i_k}$ will be referred to as
data, that of all regression parameters, i.e.\ $\beta$ and all $\beta_{i_1\dots i_d}$, as coefficients and that of
all covariance hyperparameters as covariances, e.g.\ we will use 
$p($data\,|\,covariances$)$ to refer to marginal
likelihood of observations with regression parameters integrated out.   

By allowing rank deficient covariance matrices we can fit in this
framework mixed-effects  models (also known as linear mixed models) 
as discussed for example in 
Section 13.3 of \cite{gelman:hier}.
To clarify the concept, an example more complex than \eqref{eq:model_2_levels} is helpful.
Suppose we have data on individuals
$j$ that belong to different countries $i$, that we model as a
regression $y_{ij} \sim \Nor(\alpha+\theta_i z_{ij}+\gamma_i
w_{ij},\sigma_i^2)$, where $z_{ij}$ and $w_{ij}$ are individual-level
covariates. We want to allow variation of the corresponding
coefficients that we partially explain using a country-level covariate
$s_i$, $\theta_i \sim \Nor(a+b s_i,\tau)$ and $\gamma_i \sim
\Nor(c+ds_i,\lambda)$ with $\theta_i$ and $\gamma_i$ independent. In
order to obtain a factor graph that is a tree, we need to clump
coefficients at the deepest level into one vector
$\beta_i=(\alpha_i,\theta_i,\gamma_i)^T$, and the covariates and a
vector of 1's into the
matrix $X^y_i$. The $\alpha_i$'s are artificial copies of $\alpha$. Even
though we model the variation on $\theta_i$ independently of that on
$\gamma_i$, having them clumped together forces us to work with them
as a single vector. Hence, we write $\beta_i \sim \Nor(X_i \beta,\Sigma)$
where $\beta = (\alpha,a,b,c,d)^T$, $\Sigma =
\textrm{diag}(0,\tau,\lambda)$, and $X_i$ is a $3\times 5$ matrix with
rows $(1,0,0,0,0)$, $(0,1,s_i,0,0)$, and $(0,0,0,1,s_i)$. Hence, we have obtained the nested
multilevel structure of \eqref{eq:nested_models} with $k=1$. 
An example of a structure that does not fit
the framework of \eqref{eq:nested_models} is  the crossed-effects model, 
$ y_{ij} \sim
\Nor(\alpha_i+\theta_{j},\sigma^2)$,
to which the belief propagation algorithm we describe below does not
apply.

\subsection{Message passing for posterior simulation}\label{sec:bp-nest}
We describe the messages for belief propagation in nested multilevel
models as defined in \eqref{eq:nested_models},  and how they can be
used for computing $p($data\,|\,covariances$)$ and
simulating from $p($coefficients\,|\,data,covariances$)$. The factor graph
associated with \eqref{eq:nested_models}  is a
tree;  throughout our presentation we will take $\beta$ to be its
root. The factor
nodes will be denoted by $f_{i_1\ldots i_d}$; as already discussed
the factor node that links {$y_{i_1\ldots
  i_k}$ to $\beta_{i_1\ldots i_k}$} will be denoted by
$f^y_{i_1\ldots i_k}$ and that linking {$\beta_{i_1\ldots
    i_{k}}$ to $\beta_{i_1\ldots i_{k-1}}$} will be denoted by 
$f_{i_1\ldots i_k}$; see Figure \ref{fig:fact_graph}.
Messages are functions that can be
parameterised in terms of a triplet
$(c,C,u)$, where $c$ is a positive scalar, $C$ a
positive semidefinite matrix and $u$ a vector, and for brevity we write 
\begin{equation}
{m} =  (c,C,u) \quad \textrm{to imply} \quad m(x) =
{c}\exp\big(-(1/2) x^T C x + u^Tx\big). 
\label{eq:message-generic}
\end{equation}
First we show how to update this set of parameters during the forward step, which works
from leaves to the root, i.e., from the deepest level in the hierarchy
to the most shallow. 
With every factor node, except for those  at
the deepest level linked to the observations, there are associated two
messages, one that arrives from the variable node deeper and one that
is sent to the variable node one level higher up. We distinguish
between the two corresponding triplets by using tildes for the
latter.
In this context, the message that arrives at a variable node from below coincides 
with the density of the data on the leaves that originate from the given branch of the tree,
conditional on the variable at this node but marginal with respect to 
all other variables in-between.
The messages from factor to variable nodes at the leaves are 
\begin{equation}
  \label{eq:message-leaves}
\begin{aligned}
 m_{f^y_{i_1\ldots i_k} \to \beta_{i_1\ldots i_k}} 
 &=
  (\tilde{c}_{i_1\ldots i_k}^y,\tilde{C}_{i_1\ldots i_k}^y,\tilde{u}_{i_1\ldots i_k}^y)
  \\
  \tilde{C}_{i_1\ldots i_k}^y
  &=
  (1/\sigma_{i_1\ldots i_k}^2) (X^y_{i_1\ldots i_k})^T X^y_{i_1\ldots i_k}
  \\
 \tilde{c}_{i_1\ldots i_k}^y
 &=
  (2\pi \sigma_{i_1\ldots i_k}^2)^{-d_{i_1\ldots i_k}/2}
  \exp\big(-(1/2\sigma_{i_1\ldots i_k}^2) y_{i_1\ldots i_k}^T y_{i_1\ldots i_k}\big)
  \\
  \tilde{u}_{i_1\ldots i_k}^y &= (1/\sigma_{i_1\ldots i_k}^2) (X_{i_1\ldots i_k}^y)^T y_{i_1\ldots i_k}
\end{aligned}
\end{equation}
where $d_{i_1\ldots i_k}$ is the size of $y_{i_1\ldots i_k}$. 
At higher levels, the factor to variable messages are
\begin{equation}
  \label{eq:message-factor-inner}
\begin{aligned}
  m_{f_{i_1\ldots i_d} \to \beta_{i_1\ldots i_{d-1}}} & =
  (\tilde{c}_{i_1\ldots i_d},\tilde{C}_{i_1\ldots i_d},\tilde{u}_{i_1\ldots i_d}) \\
  \tilde{c}_{i_1\ldots i_d} & = c_{i_1\ldots i_d}|G|^{1/2}
  \exp\big(-(1/2) u_{i_1\ldots i_d}^T \Gamma^T G \Gamma u_{i_1\ldots i_d} \big)  \\\tilde{C}_{i_1\ldots i_d}
  & = X_{i_1\ldots i_d}^T B^T
  (B \Sigma_{i_1\ldots i_{d-1}} B^T + I)^{-1} B  X_{i_1\ldots i_d}  
  \\
  \tilde{u}_{i_1\ldots i_d} & =  X_{i_1\ldots i_{d}}^T
  (\Sigma_{i_1\ldots i_{d-1}} C_{i_1\ldots i_{d}} + I)^{-1} u_{i_1\ldots i_{d}}
\end{aligned}
\end{equation}
where $B$, $\Gamma$ and $G$ implicitly depend on $(i_1\ldots i_{d})$ and are defined by
\begin{equation}
  \label{eq:message-factor-inner-last-line}
\begin{aligned}
C_{i_1\ldots i_{d}} = B^T B, & \quad  \Sigma_{i_1\ldots i_{d-1}} =
\Gamma^T \Gamma, \quad 
G  = (\Gamma C_{i_1\ldots i_d} \Gamma^T +I)^{-1}\,.
\end{aligned}
\end{equation}
The triplet $(c_{i_1\ldots i_d},C_{i_1\ldots i_d},u_{i_1\ldots i_d})$ in \eqref{eq:message-factor-inner} corresponds to the message $m_{\beta_{i_1\ldots i_d} \to f_{i_1\ldots i_{d}}}$.
The matrix decompositions in \eqref{eq:message-factor-inner-last-line} are used with Schur's complement to obtain
formulae above that are valid without any assumptions on invertibility
of either $C_{i_1\ldots i_d}$ or $\Sigma_{i_1\ldots i_{d-1}}$.
$\Sigma_{i_1\ldots i_{d-1}} C_{i_1\ldots i_d} + I$ is invertible by contruction
because $\Sigma_{i_1\ldots i_{d-1}}$ is positive semidefinite.

The messages from 
variable to factor nodes are as follows: at the deepest level
\begin{equation}
  \label{eq:messages-variables-to-factor-leaves}
\begin{aligned}
& m_{\beta_{i_1\ldots i_k} \to f_{i_1\ldots i_{k}}}  =
(c_{i_1\ldots i_{k}}, C_{i_1\ldots i_{k}}, 
u_{i_1\ldots i_{k}}) =  \left(\tilde{c}_{i_1\ldots i_k}^y,
    \tilde{C}_{i_1\ldots i_k}^y ,\tilde{u}_{i_1\ldots
    i_k}^y \right)
\end{aligned}
\end{equation}
while for any $d\in\{1,\dots,k-1\}$
\begin{equation}
  \label{eq:messages-variables-to-factor}
\begin{aligned}
& m_{\beta_{i_1\ldots i_d} \to f_{i_1\ldots i_{d}}}  =
(c_{i_1\ldots i_{d}}, C_{i_1\ldots i_{d}}, 
u_{i_1\ldots i_{d}}) =  \left({\prod_j} \tilde{c}_{i_1\ldots i_dj}, \sum_j 
    \tilde{C}_{i_1\ldots i_dj} ,\sum_j \tilde{u}_{i_1\ldots
    i_dj} \right)
\end{aligned}
\end{equation}
where $j$ runs over the appropriate index set, which depends on
the numbers of offsprings $\beta_{i_1\ldots i_d}$ has on the
Bayesian network. 

At the root we collect messages from  $f_i$ for $i \geq 1$, 
which come from the level below, and from $f_0$, which
comes from the prior. The normalised message is the posterior density
at the root:
\begin{equation}
\label{eq:post-root}
\begin{aligned}
\beta \mid \textrm{data}, \textrm{covariances} &  \sim \Nor(P^{-1}m,P^{-1}) \\
P = P_0 + V_0^{-1} \quad & \textrm{or} \quad P= P_0,   \quad P_0 =  \sum_i \tilde{C}_i \\
m = \sum_i \tilde{u}_i + V_0^{-1} \mu_0 \quad & \textrm{or} \quad m = \sum_i \tilde{u}_i
\end{aligned}
\end{equation}
where the alternative expressions depend on whether a Gaussian or flat
prior is used for $\beta$.
Flat prior leads to proper posterior
provided $P_0$ defined above is positive definite. Additionally,
\begin{equation}
  \label{eq:ML}
  p(\textrm{data} \mid \textrm{covariances}) = {  \Nor(x;\mu_0,V_0) \prod_i \tilde{c}_i
    \exp\left\{-(1/2) x^T \tilde{C}_i x + \tilde{u}_i^T x\right\}
    \over \Nor(x; P^{-1} m, P^{-1}) },
\end{equation}
where the first term in nominator is omitted for flat prior, and any
value $x$ can be used. Then, for any intermediate level regression
coefficients, we have 
\begin{equation}
  \label{eq:post-in}
\begin{aligned}
  \beta_{i_1\ldots i_d} \mid   & \beta_{i_1\ldots
    i_{d-1}},\textrm{data},
  \textrm{covariances} 
  \sim 
  \\
  &\Nor \left\{
 G (X_{i_1\ldots i_{d}} \beta_{i_1\ldots i_{d-1}} +
\Sigma_{i_1\ldots i_{d-1}} u_{i_1\ldots i_{d}}) ,   \Gamma^T
  (\Gamma C_{i_1\ldots i_{d}} \Gamma^T+I)^{-1} \Gamma \right\}\\ G &  =  (\Sigma_{i_1\ldots i_{d-1}}C_{i_1\ldots i_{d}}  +I)^{-1}, \quad  \Sigma_{i_1\ldots i_{d-1}} = \Gamma^T \Gamma.
\end{aligned}
\end{equation}
Simulating backwards according to the distributions described in
\eqref{eq:post-root} and \eqref{eq:post-in} we obtain a draw from
$p(\textrm{coefficients} | \textrm{data},\textrm{covariances})$. 

We have not given the details of the calculations that produce the
formulae for the messages and the conditional distributions. We have
worked under a framework mathematically richer than that of
\eqref{eq:messages}, where the definition of a factor graph is extended to
correspond to a factorisation of the joint probability measure in
terms of regular conditional densities, along the lines of the
disintegration theorem as in  Theorem 5.4 of \cite{kallenberg}. This
type of construction is suitable for the factor graph representation
of Bayesian networks with conditional Gaussian distributions with
semi-definite covariance matrices. In the context of a multilevel model, the messages are
Radon-Nikodym derivatives between Gaussian measures. It can be checked
(but we have omitted the details here) that indeed the posterior
Gaussian laws
\eqref{eq:post-in} are absolutely continuous with respect to the prior
Gaussian laws in the definition of the multilevel model in
\eqref{eq:nested_models} with Radon-Nikodym derivative proportional to
the message $m_{f_{i_1\ldots i_{d+1}}  \to \beta_{i_1\ldots i_d} }$. 

\subsection{Complexity considerations}
Both computation of $p($data\,|\,covariances$)$ and  simulation from
$p($coefficients\,|\,data,covariances$)$
via belief propagation as described above involve a computational
cost that scales linearly with the total number of regression coefficient
vectors. The forward
and backward steps
at each level involve computations that without
further structural assumptions scale cubically with the
dimension of the level-specific coefficients. 
On the other hand, if
we wish to draw several samples for the coefficients for given values
of the covariances, the cost per step can be made quadratic in the dimension of the coefficient vector, since matrix decompositions do
not have to be redone. 
 In other words, whereas the
computational cost of the algorithm has the same dependence on the
characteristics of the model as that of a Gibbs sampling algorithm
that simulates each regression coefficient conditionally on the rest, 
it achieves exact draws from the high-dimensional distribution $p($coefficients\,|\,data,covariances$)$.
On the other hand, to obtain a factor tree graph we might have to clump
together variables, hence increasing the dimension of the variable node
dimensions, as we did for the mixed-effects model example in Section \ref{sec:model_formulation}.
In such cases the computational cost per iteration of the Gibbs sampler will be smaller. However, such differences will typically be small
relative to the overall cost of the algorithms. Additionally, one can
use belief propagation to update a subset of the variables conditionally on the rest, as we discuss in the next section, hence obtaining again the same cost per iteration as the Gibbs sampler.
Sampling using belief propagation lends itself to parallelisation, but we do not develop
this idea further here, although it should be considered for software development.

\section{Blocked Gibbs sampling and marginal algorithms}

The algorithm of Section \ref{sec:bp-nested} can be used in the popular context where the covariance components are given a prior distribution and Bayesian inferences are performed on the joint distribution $p(\textrm{coefficients},\textrm{covariances}\,|\, \textrm{data})$.
Posterior computations are typically performed using a Gibbs sampling scheme that alternates sampling from $p($covariances\,|\,data, coefficients$)$ and $p(\textrm{coefficients}|\textrm{covariances},\textrm{data})$.
If conditionally conjugate priors (e.g. Wishart) are used, sampling from $p(\textrm{covariances}|\textrm{data},\textrm{coefficients})$ can be efficiently accomplished exploiting the conditional independence across
covariance components given the coefficients and the data.
Belief propagation can then be used to sample from $p($coefficients\,|\,data,covariances$)$ efficiently.

An alternative approach is to perform inferences on the marginal space $p($covariances\,| data$)$.
Even in this case, the belief propagation algorithm of Section \ref{sec:bp-nested} is crucial to allow efficient point-wise evaluation of the marginal likelihood $p($data\,|\,covariances$)$, and thus of $p($covariances\,|\,data$)$ up to proportionality.
The relative merits of working with the extended posterior $p($co\-ef\-fi\-cients,\,covariances\,|\,data$)$ or with the marginal one $p($covariances\,| data$)$ are case-specific. 
In general one would expect the marginal approach to be preferable when the distribution $p(\textrm{covariances}| \textrm{data})$ is small or moderate dimensional, so that numerical or Monte Carlo integration can be efficiently employed in the marginal space. 
When instead the distribution $p(\textrm{covariances}\,|\, \textrm{data})$ is high-dimensional it may be challenging to perform inferences in the marginal space because, having integrated out the regression coefficients, the covariance components are not anymore conditionally independent and one ends up with a potentially complex high-dimensional correlated distribution. In this case the blocked Gibbs Sampling approach may be more efficient.

The methodology we presented so far is well-suited to contexts
where the number of regression coefficients is large (i.e.\ tree with many nodes) and the size of each coefficient is small,
since the cost of operations grows as the cube of that size but only
linearly with the number of nodes in the tree.
When the size of variable nodes is large, we can
still use the message passing algorithm of Section \ref{sec:bp-nested}
to update a subset of regression coefficients (i.e. some components of each $\beta_{i_1\dots i_d}$) jointly across all nodes of
the tree conditionally on the remaining coefficients. 
The resulting blocked Gibbs sampling scheme can still exhibit large improvements over the single-site updating.
A more detailed study of this case is left to future work.

\section{Synthesis of the related literature}

The forward-backward iteration we describe for sampling efficiently
the high-dimensional Gaussian distribution, $p(\textrm{coefficients} |
\textrm{data},\textrm{covariances})$, is a generalisation of the
so-called forward filtering backward sampling algorithm, which is
used in the time series community for Bayesian inference for state
space models, see for example Algorithm 13.4 in \cite{sylvia}. State
space models have factor graphs with single-branch tree structure.
The extension of Kalman filtering recursion to tree structures
has been long-known, especially in the context of multiscale systems (e.g.\ \citealp{Chou1994}).
Similar ideas have been exploited in spatial statistics contexts, see for example \cite{cressie} where algorithms for spatial Gaussian models with tree-structured dependence are developed.
\cite{nips} exploit Kalman filter recursions to perform posterior maximization for some multilevel models that are subset of the framework we consider here. 
These previous works typically focus on computing marginal
distributions and do not develop sampling algorithms,
and they do not make a clear connection with belief propagation.
The connection between belief propagation and Kalman recursions is
recognised in the early technical report \cite{dempster} and using a
message-passing formulation in \cite{normand}, who provide
references to works even older than \cite{dempster}. 

The role of belief propagation within Bayesian computation for
multilevel models is fully recognised in \cite{wilki02} who work along
the same lines we have followed in this note, in particular their
Section 2.4 on message passing for sampling posteriors that arise in
Gaussian tree models (the approach we use in Section \ref{sec:bp-nest}
corresponds to what they call the canonical parameterisation of the
multivariate Gaussian). Relative to that work the main novelty in this
note is that we have worked out messages with no assumptions on
invertibility of prior covariance matrices. This extension has allowed us to
also cast mixed effect models as nested multilevel models and use
belief propagation for those too. For the two-level mixed effect models
\cite{carlin99} derive an efficient algorithm for sampling the
posterior distribution of the regression coefficients, which is an
instance of the generic algorithm of Section \ref{sec:bp-nested},
where the structure of 0's in prior covariances that results from
clumping is explicitly
exploited to simplify some of the matrix computations. 

Seen as latent Gaussian models, the nested multilevel models we consider give rise to large
sparse precision matrices. \cite{wilki02} and \cite{wilki04} show how
to compute the canonical parameters of the Gaussian prior and
posterior using sparse linear algebra computations and exploiting the
graphical model structure to identify the zeros in the precision
matrices.  The potential of sparse linear algebra computations for
inference and simulation in latent Gaussian models has been recognised
at least since \cite{rue01}; this work, and inter alia the follow-ups
\cite{block} and \cite{rue:book} have illustrated how the precision
matrices that arise in spatiotemporal latent Gaussian models can be
treated using sparse linear algebra algorithms, such as algorithms for
banded matrices and algorithms that attempt to produce Choleksy
factors nearly as sparse as the posterior precision. This work as has
also  recognised the connections between their fast linear algebra
approach and Kalman filters/smoothers; see for example the overview in
Section 1.2.1 of  \cite{rue:book}. The way numerical analysis
techniques for sparse matrices can be used for computations with
Gaussian graphical models is exposed in Section 2.4 of
\cite{rue:book}, and Section 2.5 of the same book carries out a
simulation study that tries two black-box algorithms for sparse
matrices in the context of precision matrices that arise in
spatiotemporal graphical models. \cite{wilki04} suggests the use of such sparse
linear algebra methods for inference in Gaussian hierarchical models;
first compile the sparse precision using computations that exploit the
known positions of the zeros and then feed the precision into a sparse
linear algebra algorithm that returns its Cholesky factor. The article
does not study whether the structure of zeros in the precision from
common families of hierarchical models, such as for example the nested
hierarchical models considered in this note, is such that the extraction of the
Cholesky factor  is
actually fast or whether the resultant factor is sparse, or the complexity of the associated
operations. These type of questions have been studied in more depth in
the case of spatial models in Chapter 2 of
\cite{rue:book}. 

In the opposite direction, the linear algebra
community has explored the use of belief propagation as an efficient
iterative method for solving linear systems. An example of this line
of work is  \cite{bp-algebra}, who use belief propagation for solving
linear systems with positive definite matrices by linking the system solution
to the marginal means (or the mode) of a Gaussian distribution with
precision matrix given by the system matrix. They discuss connections
of this approach to direct and iterative methods for solving systems
and find it competitive. 

\section*{Acknowledgements}

The authors are grateful to Darren Wilkinson for his input.

\bibliographystyle{apalike}
\bibliography{scalable}

\end{document}